\documentclass[showpacs,preprintnumbers,prd,nofootinbib,floats,amssymb,floatfix]{revtex4-2}
\usepackage{graphicx}
\usepackage{amsxtra}
\usepackage{hyperref}
\usepackage{amssymb}
\usepackage{amstext}
\usepackage{amsmath}
\usepackage{cleveref}
\usepackage{stackrel}
\usepackage{blkarray}
\begin{document}
\title{ Canonical analysis of linearized $\lambda R$ gravity plus a  Chern-Simons term }
\author{Alberto Escalante}  \email{aescalan@ifuap.buap.mx}
\author{Jes\'us Aldair Pantoja-Gonz\'alez }\email{jpantoja@ifuap.buap.mx}
\author{Victor Julian P\'erez-Aquino}  \email{victor.perezaq@alumno.buap.mx}
 \affiliation{Instituto de F\'isica, Benem\'erita Universidad Aut\'onoma de Puebla. \\ Apartado Postal J-48 72570, Puebla Pue., M\'exico, }

\begin{abstract}
The Hamiltonian analysis for the linearized  $\lambda R$ gravity plus a Chern-Simons term is performed. The first-class and second-class constraints for arbitrary values of $\lambda$ are presented, and one physical degree of freedom is reported. The second-class constraints are removed, and the corresponding generalized Dirac brackets are constructed; then, the difference between theories with different values of $\lambda$  is remarked.
\end{abstract}
 \date{\today}
\preprint{}
\maketitle

\section{Introduction}

In the pursuit of constructing a consistent quantum theory of gravity, some interest has been gained in the study of higher-order modified gravity models. In these models, one deals with different kinds of extensions or deformations of General Relativity (GR) by including higher-order terms of the curvature in the action \cite{Weyl,Chen}. Including such extra terms can be considered an ``improvement" of GR because of the renormalizability of the resulting theory \cite{Stelle}. However, a not-so-attractive feature of the higher-order theories is the existence of Ostrogradsky's ghosts \cite{Ostrogradsky}. In fact, in the modified gravity context, an Ostrogradsky ghost is a massive or not massive particle with negative energy; such an extra degree of freedom leads to unitarity problems (see \cite{Woodard} for a comprehensive exposition). The existence of an  Ostrogradsky mode is a direct consequence of the inclusion of higher-order time derivatives of the metric in the action, but, in principle, one can avoid this pathology by degeneracy arguments \cite{Ganz}. Another way to get rid of ghosts is due to Ho\v{r}ava and his quantum field theory of gravity; see \cite{Horava1} for the original work and \cite{Herrero} for the actual status of the theory.\\
Ho\v{r}ava's proposal is an extension of the ADM formalism of GR where the violation of Lorentz symmetry (in the high energies limit) and the anisotropic scaling symmetry are the key ingredients, among other peculiarities. Such violation allows us to consider additional terms of the curvature that contain only higher-order spatial derivatives of the three-metric. The requirement to include the extra terms is that they must be invariant under foliation-preserving diffeomorphisms.\\
Despite the technical difficulties of Ho\v{r}ava's gravity, the resulting theory is,  in principle, renormalizable and free of ghosts \cite{Barvinsky}. Such difficulties are evident when we consider the number of permissible additional terms. Nevertheless, it is possible to reduce the number of extra terms if one proposes that the lapse function depends only on time $N=N(t)$; this eventually allows one to select $N=1$ because of the reparametrization invariance. This reduction is known as projectable Ho\v{r}ava gravity. On the other hand, one can retain a general dependence of the lapse $N=N(t,x)$ and truncate the potential. Furthermore, the short action of the  Ho\v{r}ava model is characterized by some parameter $\lambda$, called the  $\lambda R$ gravity; it is a slight deviation from GR. There are two cases under study when $\lambda \neq \frac{1}{2}$, and when $\lambda =\frac{1}{2}$ the so-called critical value; when $\lambda=1$ Einstein's theory is recovered; for deeper insight on the role of the parameter $\lambda$ see \cite{Loll,Ambjorn}. The $\lambda R$ model has been studied in order to give some theoretical support to the full Ho\v{r}ava's theory from a Hamiltonian point of view \cite{Bellorin}.\\
Now, as said above, the Ho\v{r}ava theory admits spatial higher-order terms, but even so, it is still interesting to consider ``fully" (spatial, temporal) higher-order terms. In fact, it is known that by adding, for example, a Chern-Simons term to a particular case of Ho\v{r}ava's theory, it is possible to construct chiral gravitational waves \cite{Myung}. \\
By following the above  line of thought, in this paper, we will consider the coupling between the higher order Chern-Simons action and the $\lambda R$ model in a perturbative context by taking an expansion around the Minkowski background. In fact, we will not work with a perturbative ADM formalism as it is usually reported in the literature. We will work with the usual Fierz-Pauli Lagrangian, and by introducing a set of variables, it will be written as a Ho\v{r}ava's-like theory. Thus, the canonical study will be direct and very suitable because of its drivability but non-triviality. We believe that the study of this kind of theory is an excellent laboratory for testing classical and quantum ideas of gauge systems supported by the Hamiltonian analysis. We need to remember that the canonical analysis is mandatory for the study of future quantum aspects. In this regard,  we can gain some insights into this kind of theory by completely controlling the constraints and the canonical structure; for example, we will show that the contribution of the  $\lambda \neq \frac{1}{2}$ parameter is relevant in the final structure of the Dirac brackets. On the other hand, for the case $\lambda=\frac{1}{2}$ the constraints and the fundamental Dirac's brackets are different from those for  $\lambda \neq \frac{1}{2}$. Thus, adding the Chern-Simons term completely modifies the canonical structure of  $\lambda R$ gravity.\\
The paper is organized as follows. Section II presents the fundamental aspects of $\lambda R$ gravity plus a Chern-Simons term. In subsections A and B, we develop a Hamiltonian analysis for the cases $\lambda \neq \frac{1}{2}$ and $\lambda = \frac{1}{2}$ respectively. Section II is regarded to conclusions.\\

\section{Linearized $\lambda R$ gravity plus a Chern-Simons term}
It is well-known that linearized gravity is described by  the Fierz-Pauli Lagrangian for massless particles of spin two,   given by 
\begin{equation}\label{grav lin}
    \mathcal{L}_{EH} = \frac{1}{4} \partial_\lambda h_{\mu \nu} \partial^\lambda h^{\mu \nu} - \frac{1}{4} \partial_\lambda h^\mu _\mu \partial^\lambda h^\nu _\nu + \frac{1}{2} \partial_\lambda h^\lambda _\mu \partial^\mu h^\nu _\nu - \frac{1}{2} \partial_\lambda h^\lambda _\mu \partial_\nu h^{\nu \mu},
\end{equation}
where  $h_{\mu \nu}$ is a perturbation around the Minkowski background with metric $\eta_{\mu \nu}=(-1, 1, 1)$. This Lagrangian can be easily deduced from the perturbation of the Einstein-Hilbert (EH) action and it propagates two degrees of freedom. On the other hand, the perturbative version of the Chern-Simons action (CS) is expressed as
\begin{equation}\label{lag CS lin}
    \mathcal{L}_{CS} = \frac{\epsilon^{\lambda \mu \nu}}{2\mu} \left(\frac{}{} \partial_\sigma h_\lambda ^\rho \partial_\rho \partial_\mu h^\sigma _\nu - \partial_\sigma h^\rho _\lambda \partial^\sigma \partial _\mu h_{\rho \nu} \frac{}{}\right), 
\end{equation}
this action does not propagate degrees of freedom and it is a  higher-order theory in the time derivatives as we will show below. By adding   (\ref{lag CS lin})  into (\ref{grav lin})  and performing the $2+1$ decomposition we obtain the well-known expression for topological massive gravity
\begin{align}
        \mathcal{L} & = \frac{1}{4} (\dot{h}_{ij}\dot{h}^{ij}-\dot{h}_i ^i \dot{h}_j ^j) + (\partial^i h_j ^j - \partial_j h^{ij})\dot{h}_i ^0 - \frac{1}{2} \partial_i h_{0j} \partial^i h^{0j} - \frac{1}{4} \partial _i h_{jk} \partial^i h^{jk} \nonumber \\
        & + \frac{1}{2} \partial_i h^{00}\partial^ih^j _j + \frac{1}{4}\partial_i h^j _j \partial^i h^k _k - \frac{1}{2}\partial_i h^i _j \partial^j h^{00} - \frac{1}{2} \partial _i h^i _j \partial^j h^k _k + \frac{1}{2}\partial _i h^i _0 \partial_j h^{j0} + \frac{1}{2}\partial_i h^i _j \partial_k h^{kj}\nonumber \\ 
        & + \frac{\epsilon^{ij}}{\mu} \left(\frac{}{} \dot{h}^k _0 \partial_k \partial_i h_{0j} - \partial_i \partial _k h_{00} \dot{h}^k _j -\partial_i h^k _0 \ddot{h}_{kj} + \ddot{h}_i ^k \partial_k h_{0j} + \frac{1}{2} \partial_k \partial_l h^l _i \dot{h}^k _j + \frac{1}{2} \dot{h}^k _i \ddot{h}_{kj}\right. \nonumber \\
        & \left. + \frac{1}{2} \nabla^2 h^0 _i \dot{h}_{0j} + \frac{1}{2} \nabla^2 h^k _i \dot{h}_{kj} +  \partial_k h^l _0 \partial_i \partial_l h^k _j - \nabla^2 h^k _0 \partial_i h_{kj} - \nabla^2 h_{00} \partial_i h_{0j} \right),   
\end{align}
we can observe that this action presents second order time derivatives; its study has been reported in \cite{Barcelos, Esca} by using the Dirac  and the GLT frameworks respectively. For our purposes, we will introduce the following set of variables
\begin{equation}{\label{variable K}}
    K_{ij} = \frac{1}{2} \left( \dot{h}_{ij} - \partial_ih_{0j} - \partial_j h_{0i} \right), 
\end{equation}
thus the action takes the following form
\begin{align}\label{cap3: L_EH-CS K}
        \mathcal{L} & = K_{ij} K^{ij} - K^i{}_i K^j{}_j - \frac{1}{2} h_{00} R_{ij}{}^{ij} - \frac{1}{2} h^{ij} \left(\frac{}{} R_{ikj}{}^k - \frac{1}{2}\delta_{ij} R_{lm}{}^{lm} \frac{}{}\right)\nonumber\\ 
        & + \frac{\epsilon ^{ij}}{\mu} \left( 2K^k_j \dot{K}_{ki} - 2K_{ik} \partial_j \partial^k h_{00} - K^k_i \partial_k \partial_l h^l_j + K_{ki} \nabla^2 h^k_j \right. \nonumber \\ 
        & \left. + \frac{1}{2} h^k_0 \partial_i \nabla^2 h_{kj} -\frac{1}{2} h^k_0 \partial_k \partial _l \partial_i h^l _j \right) + \alpha^{ij} \left( \dot{h}_{ij} - 2\partial_ih_{0j} - 2 K_{ij} \right),
\end{align}
where  $\alpha^{ij}$ are Lagrange multipliers enforcing (\ref{variable K}). Now we can rewrite the action in a  H\v{o}rava-like fashion by introducing the generalized De Witt metric
\begin{equation}
    G^{ijkl}=\frac{1}{2}\left(\delta^{ik}\delta^{jl}+\delta^{il}\delta^{jk}\right)-\lambda\delta^{ij}\delta^{kl},
\end{equation}
with its corresponding inverse given by
\begin{equation}
\mathcal{G}_{ijkl}=\frac{1}{2}\left(\delta_{ik}\delta_{jl}+\delta_{il}\delta_{jk}\right)+\frac{\lambda}{1-2\lambda}\delta_{ij}\delta_{kl},
\end{equation}
hence, the Lagrangian  takes the following new fashion 
\begin{align}\label{L1}
        \mathcal{L} & = G^{ijkl} K_{ij} K_{kl} - \frac{1}{2} h_{00} R_{ij}{}^{ij} - \frac{1}{2} h^{ij} \left(\frac{}{} R_{ikj}{}^k - \frac{1}{2}\delta_{ij} R_{lm}{}^{lm} \frac{}{}\right)\nonumber\\
        & + \frac{\epsilon ^{ij}}{\mu} \left( 2K^k_j \dot{K}_{ki} - 2K_{ik} \partial_j \partial^k h_{00} - K^k_i \partial_k \partial_l h_j^l + K_{ki} \nabla^2 h^k_j \right.\nonumber\\
        & \left. + \frac{1}{2} h^k_0 \partial_i \nabla^2 h_{kj} -\frac{1}{2} h^k_0 \partial_k \partial _l \partial_i h^l _j \right) + \alpha^{ij} \left( \dot{h}_{ij} - 2\partial_ih_{0j} - 2 K_{ij} \right).
\end{align}
It is worth commenting that the definition of $K_{ij}$ does not imply that it is a dynamical variable, but since there are higher-order time derivatives in the EH+CS action, we can take it as such, this fact will reduce the order of the derivatives and the canonical analysis can be performed directly. At the end of the study, we will recognize the $h_{\mu\nu}$ field as the actual dynamical field. However, at this point, we effectively write the total action as a $\lambda R$ model; at the end, we are deviating from linearized gravity, and such deviation is codified in the parameters $\mu$ and $\lambda$. If we let $\mu\rightarrow\infty$ and $\lambda=1$ we recover linearized gravity. From now on, we will perform the canonical analysis for two separate cases: one for $\lambda\neq\frac{1}{2}$ and the other one for $\lambda=\frac{1}{2}$ due to a singular value in the inverse of the generalized De Witt metric when $\lambda=\frac{1}{2}$.

\subsection{Canonical Analysis for $\lambda \neq \frac{1}{2}$}

We start by calculating the canonical momenta of the action (\ref{L1}), given by
\begin{equation}
\label{CM1}
    \begin{split}
        \pi^{00} & = \frac{\partial \mathcal{L}}{\partial \dot{h}_{00}} = 0,\\
        \pi^{0i} & = \frac{\partial \mathcal{L}}{\partial \dot{h}_{0i}} = 0,\\
        \pi^{ij} & = \frac{\partial \mathcal{L}}{\partial \dot{h}_{ij}} = \alpha^{ij},\\
        \tau^{ij} & = \frac{\partial \mathcal{L}}{\partial \dot{\alpha}_{ij}} = 0,\\
        P^{ij} & = \frac{\partial \mathcal{L}}{\partial \dot{K}_{ij}} = \frac{1}{\mu}\left( \epsilon^{ik}K^j_k + \epsilon^{jk}K^i_k \right).
    \end{split}
\end{equation}
By using the canonical momenta, we arrive directly to the canonical Hamiltonian
\begin{align}
    \mathcal{H}_{C} & = -G^{ijkl}K_{ij}K_{kl} + \frac{1}{2}h_{00} R_{ij}{}^{ij}+ \frac{1}{2}h^{ij}\left( \frac{}{}R_{ikj}{}^{k} - \frac{1}{2}\delta_{ij}R_{lm}{}^{lm} \frac{}{}\right)\nonumber\\
    & + \frac{\epsilon^{ij}}{\mu} \left(\frac{}{} 2K_{ki}\partial_j \partial^kh_{00} +K^k_i\partial_k\partial_lh^l_j-K_{ki}\nabla^2h^k_j\right.\nonumber\\ 
    & \left. -\frac{1}{2}h^k_0\partial_i\nabla^2h_{kj} + \frac{1}{2}h^k_0\partial_k\partial_l\partial_ih^l_j \frac{}{}\right) + 2\pi^{ij}\left(\frac{}{} \partial_ih_{0j}+K_{ij}\frac{}{}\right).
\end{align}
The nature of the momenta indicate us that we are dealing with a constrained system, the primary constraints are the following:
\begin{equation}
\label{PrimaryConstraints1}
    \begin{split}
        \phi^{00} & : \pi^{00} \approx 0,\\
        \phi^{0i} & : \pi^{0i} \approx 0,\\
        \phi^{ij} & : \pi^{ij} - \alpha^{ij} \approx 0,\\
        \psi^{ij} & : \tau^{ij} \approx 0,\\
        \Sigma^{ij} & : P^{ij} - \frac{1}{\mu}\left( \epsilon^{ik}K^j_k + \epsilon^{jk}K^i_k \right) \approx 0,\\
        \Sigma & : P \approx 0,
    \end{split}
\end{equation}
the correspondent primary Hamiltonian takes the form 
\begin{equation}
\mathcal{H}_{1} = \mathcal{H}_{C} + \lambda_{\mu \nu} \phi^{\mu \nu} + \zeta_{ij} \psi^{ij} + \xi_{ij}\Sigma^{ij},
\end{equation}
where $\lambda_{\mu\nu}$, $\zeta_{ij}$ y $\xi_{ij}$ are Lagrange multipliers enforcing the primary constraints. From consistency of the primary constraints we obtain the following secondary constraints
\begin{equation}
    \begin{split}
        & S = \frac{1}{2}R_{ij} {}^{ij} +  \frac{2}{\mu}\epsilon^{ij}\partial_j\partial^kK_{ik} \approx0,   \label{S}\\ 
        & S^{0i}  = \frac{\epsilon^{jk}}{4\mu} \partial_j \nabla^2 h^i_k - \frac{\epsilon^{jk}}{4\mu} \partial^i\partial_l\partial_jh^l_k + \partial_j\pi^{ij} \approx 0,\\
        & W = 2(1-2 \lambda )K - \frac{\epsilon^{ij}}{\mu}\partial_i\partial_lh^l_j - 2\pi \approx 0.
    \end{split}
\end{equation}
We can observe a contribution of $\lambda$ in the constraint $W$. We did not find further constraints from the evolution of secondary constraints in time. In this manner, the complete set of constraints is given by
\begin{equation}
    \begin{split}
        & \phi^{00} : \pi^{00} \approx 0,\\
        & \phi^{0i} : \pi^{0i} \approx 0,\\
        & \phi^{ij} : \pi^{ij} - \alpha^{ij} \approx 0,\\
        & \psi^{ij} : \tau^{ij} \approx 0,\\
        & \Sigma^{11} : P^{11} - \frac{2}{\mu}K^1_2 \approx 0,\\
        & \Sigma^{12} : P^{12}-\frac{1}{\mu}(K^2_2-K^1_1) \approx 0,\\
        & \Sigma : P \approx 0,\\
        & S : \frac{1}{2}R_{ij}{}^{ij} + \frac{2}{\mu}\epsilon^{ij}\partial_j\partial^kK_{ik} \approx 0, \\
        & S^{0i} : \partial_j\pi^{ij} + \frac{\epsilon^{jk}}{4\mu}( \partial_j \nabla^2 h^i_k - \partial^{i}\partial_{l}\partial_{j}h^{l}_{k}) - \frac{\epsilon^{jk}}{4\mu} (\partial_j\nabla^{2}\tau^i_k - \partial^{i}\partial_{l}\partial_{j}\tau^{l}_{k}) \approx 0 \\
        & W :  2(1-2\lambda)K - \frac{1}{\mu} \epsilon^{ij}\partial_i\partial_lh^l_j - 2\pi \approx 0.
    \end{split}
\end{equation}
Since we have no more constraints, we can proceed with the classification into first-class and second-class constraints.  For this aim, we calculate the nonzero Poisson brackets between all constraints   given by 
\begin{eqnarray}
\lbrace \phi^{ij},\psi^{kl} \rbrace &=& - \frac{1}{2}(\delta^{ik}\delta^{jl}+\delta^{il}\delta^{jk}) \delta^{2}(\vec{x}-\vec{y}),
\\
\lbrace \phi^{ij},S \rbrace &=& - \frac{1}{2}(\partial^{i}\partial^{j} - \delta^{ij}\nabla^{2}) \delta^{2}(\vec{x}-\vec{y}),
\\
\lbrace \phi^{ij},S^{0k} \rbrace &=& \frac{1}{8\mu}[\epsilon^{il}(\partial^{j}\partial^{k}+\delta^{jk}\nabla^{2}) + \epsilon^{jl}(\partial^{i}\partial^{k}+\delta^{ik}\nabla^{2})]\partial_{l} \delta^{2}(\vec{x}-\vec{y}),
\\
\lbrace \phi^{ij},W \rbrace &=& - \frac{1}{2\mu}(\epsilon^{ik}\partial^{j} + \epsilon^{jk}\partial^{i})\partial_{k} \delta^{2}(\vec{x}-\vec{y}),
\\
\lbrace \Sigma^{ij},\Sigma^{kl} \rbrace &=& - \frac{1}{\mu}(\epsilon^{ik}\delta^{jl} + \epsilon^{jk}\delta^{il} + \epsilon^{il}\delta^{jk} + \epsilon^{jl}\delta^{ik}) \delta^{2}(\vec{x}-\vec{y}),
\\
\lbrace \Sigma^{ij},S \rbrace &=& \frac{1}{\mu}(\epsilon^{ik}\partial^{j} + \epsilon^{jk}\partial^{i})\partial_{k} \delta^{2}(\vec{x}-\vec{y}),
\\
\lbrace \Sigma^{ij},W \rbrace &=& - 2(1-2\lambda)\delta^{ij}\delta^{2}(\vec{x}-\vec{y}),
\\
\lbrace \Sigma,W \rbrace &=& - 4(1-2\lambda)\delta^{2}(\vec{x}-\vec{y}),
\\
\lbrace S,W \rbrace &=& \frac{1}{2}\nabla^{2}\delta^{2}(\vec{x}-\vec{y}).
\end{eqnarray}
These brackets let us identify the following 6 first-class constraints:
\begin{eqnarray}\label{EH-CS: restricciones 1 class}
         &\Gamma_1& : \pi^{00} \approx 0, \nonumber \\
        & \Gamma_2^{i} &: \pi^{0i} \approx 0, \nonumber \\
        & \Gamma_3 &: \frac{1}{2} R_{ij}{}^{ij} + \frac{2}{\mu} \epsilon^{ij} \partial_j \partial^kK_{ik}+\frac{\nabla^2P}{4(1-2\lambda)} +\frac{\partial_1 \partial^1P}{4} - \frac{\partial_2 \partial^2P}{4} - \partial_1\partial^2[P^{12}-\frac{1}{\mu}(K^2_2-K^1_1) ] \nonumber \\
        &&  - \frac{ \partial_1 \partial^1}{2} (P^{11} - \frac{2}{\mu}K^1_2)+ \frac{ \partial_2\partial^2}{2} (P^{11} - \frac{2}{\mu}K^1_2)    -     \frac{1}{2} (\partial^i\partial^j - \delta^{ij}\nabla^2)\tau_{ij} \approx 0, \nonumber \\ 
        & \Gamma_4^{i}& : \partial_j\pi^{ij} + \frac{\epsilon^{jk}}{4\mu} \partial_j \nabla^2 h^i_k - \frac{\epsilon^{jk}}{4\mu} \partial^i\partial_l\partial_jh^l_k  
        - \frac{\epsilon^{jk}}{4\mu}\partial_j\nabla^2\tau^i_k + \frac{\epsilon^{jk}}{4\mu}\partial^i\partial_l\partial_j\tau^l_k \approx 0,
\end{eqnarray}
and the following 10 second-class constraints
\begin{equation}\label{EH-CS: restricciones 2 class1}
    \begin{split}
        & \chi_1^{ij} : \pi^{ij} - \alpha^{ij} \approx 0,\\
        & \chi_2^{ij} : \tau^{ij} \approx 0,\\
        & \chi_3 : P^{11} - \frac{2}{\mu}K^1_2 \approx 0,\\
        & \chi_4 : P^{12}-\frac{1}{\mu}(K^2_2-K^1_1) \approx 0,\\
        & \chi_5 : P \approx 0,\\
        & \chi_6 : 2(1-2\lambda)K - \frac{\epsilon^{ij}}{\mu} \partial_i \partial_l h^l_j - 2\pi \approx 0.
    \end{split}
\end{equation}
As mentioned before, there is a contribution of the parameter $\lambda$ in the structure of the constraints, more precisely, in the constraint $\chi_{6}$. This fact tells us that we could have a contribution of the parameter $\lambda$  to the Dirac brackets because the second-class constraints are helpful for the construction of these brackets.
\begin{equation}\label{Bdir}
    \left\lbrace A, B\right\rbrace_{D}= \left\lbrace A, B\right\rbrace-\int dudv\left\lbrace A, \chi_{a}(u)\right\rbrace C^{ab}\left\lbrace \chi_{b}(v), B\right\rbrace,
\end{equation}
where  $C_{ab}=\left\lbrace \chi_{a}, \chi_{b}\right\rbrace$ is the matrix whose entries are the Poisson brackets between all second-class constraints and its inverse  is given by   $C^{ab}$.  Therefore, we just need to calculate all the Poisson brackets between the second class constraints $\chi_{a}$, the non-zero brackets are:
\begin{eqnarray}
\{ \chi_1^{ij}(x), \chi_{2kl}(y)  \} &=& -\frac{1}{2}\delta^{ij}_{kl}\delta^2(x-y), \nonumber \\
\{ \chi_3(x), \chi_{4}(y)  \} &=& -\frac{1}{\mu} \delta^2(x-y), \nonumber \\
\{ \chi_5(x), \chi_{6}(y)  \} &=& -2(1-2\lambda) \delta^2(x-y), 
\end{eqnarray}
where $\delta_{kl}^{ij}=\delta^{i}_{k}\delta^{j}_{l}+\delta^{i}_{l}\delta^{j}_{k}$. Hence, by directly using \eqref{Bdir} we find the non-trivial Dirac's brackets: 
\begin{eqnarray}
\label{DB1}
\{ h_{ij}(x), \pi^{kl}(y) \}_D&=& \frac{1}{2}\delta^{kl}_{ij}\delta^2(x-y), \nonumber \\
\{ K_{ij}(x), P^{kl}(y) \}_D&=&\frac{1}{4} \left( \delta^{kl}_{ij}- \delta_{ij}\delta^{kl}  \right)\delta^2(x-y), \nonumber \\
\{ K_{ij}(x), \pi^{kl}(y) \}_D&=& \frac{\delta_{ij}}{4\mu (1-2\lambda)} \left(\epsilon^{mk} \partial_m \partial^l + \epsilon^{ml} \partial_m \partial^k  \right)\delta^2(x-y), \nonumber \\
\{ h_{ij}(x), K^{kl}(y) \}_D&=& \frac{\delta_{ij} \delta^{kl}}{(1-2\lambda)} \delta^2(x-y), \nonumber \\
\{ K_{ij}(x), K^{kl}(y) \}_D&=&\frac{\mu}{8} \left( \delta^{11}_{ij}\delta^{kl}_{12}  - \delta^{12}_{ij}\delta^{kl}_{11} \right)\delta^2(x-y)- \frac{\mu}{8} \left( \delta_{ij}\delta^{kl}_{12}  - \delta^{12}_{ij}\delta^{kl} \right)\delta^2(x-y), \nonumber \\
\{ P_{ij}(x), P^{kl}(y) \}_D&=&\frac{1}{4\mu} \delta^{12}_{ij} \left( \delta^{kl}_{22}  - \delta^{kl}_{11} \right) - \frac{1}{4\mu} \delta^{kl}_{12} \left( \delta^{22}_{ij}  - \delta^{11}_{ij} \right)\delta^2(x-y), \nonumber \\
\{ h_{00}(x), \pi^{00}(y) \}_D&=& \delta^2 (x-y), \nonumber \\
\{ h_{0i}(x), \pi^{0j}(y) \}_D&=& \frac{1}{2} \delta^j_i \delta^2(x-y),
\end{eqnarray}
the remaining brackets vanish. As we can note, the presence of $\lambda$ persist, now it appears in the brackets structure. We can say more about these brackets, if we take $\lambda=1$ then these brackets are reduced to those reported in \cite{Barcelos}. On the other hand, the counting of physical degrees of freedom is calculated as follows; there are 24 canonical variables, 6 first-class constraints and 10 second-class constraints, therefore we have
\begin{equation}
    DOF = \frac{1}{2}\left( 24 - 2*6-10 \right) = 1,
\end{equation}
just like topologically massive gravity, however, in this model, the $\lambda$ parameter is relevant in the final structure of the brackets. In the following section we will perform the canonical analysis at the critical point, namely for $\lambda = \frac{1}{2}$, and the differences will be discussed. 
\subsection{Canonical analysis for $\lambda = \frac{1}{2}$}
Now we will considering $\lambda=\frac{1}{2}$. The  generalized De Witt metric takes the simple form
\begin{equation}
    \hat{G}^{ijkl} = \frac{1}{2} (\delta^{ik} \delta^{jl} + \delta^{il} \delta^{jk}) -\frac{1}{2} \delta^{ij} \delta^{kl},
\end{equation}
thus, the Lagrangian (\ref{L1})  is written as 
\begin{align}
        \mathcal{L} & = K_{ij}K^{ij}- \frac{1}{2} K^2 - \frac{1}{2} h_{00} R_{ij}{}^{ij} - \frac{1}{2} h^{ij} \left(\frac{}{} R_{ikj}{}^k - \frac{1}{2}\delta_{ij} R_{lm}{}^{lm} \frac{}{}\right)\nonumber\\
        & + \frac{\epsilon ^{ij}}{\mu} \left( 2K^k_j \dot{K}_{ki} - 2K_{ik} \partial_j \partial^k h_{00} - K^k_i \partial_k \partial_l h^l_j + K_{ki} \nabla^2 h^k_j \right.\nonumber\\
        & \left. + \frac{1}{2} h^k_0 \partial_i \nabla^2 h_{kj} -\frac{1}{2} h^k_0 \partial_k \partial _l \partial_i h^l _j \right) + \alpha^{ij} \left( \dot{h}_{ij} - 2\partial_ih_{0j} - 2 K_{ij} \right).
\end{align}
The canonical momenta are given by equation \eqref{CM1}
but the canonical Hamiltonian now takes the form
\begin{align}
    \mathcal{H}_{C} & = -\frac{1}{2}K^2 -K_{ij}K^{ij}+ \frac{1}{2}h_{00} R_{ij}{}^{ij}+ \frac{1}{2}h^{ij}\left( \frac{}{}R_{ikj}{}^{k} - \frac{1}{2}\delta_{ij}R_{lm}{}^{lm} \frac{}{}\right)\nonumber\\
    & + \frac{\epsilon^{ij}}{\mu} \left(\frac{}{} 2K_{ki}\partial_j \partial^kh_{00} +K^k_i\partial_k\partial_lh^l_j-K_{ki}\nabla^2h^k_j\right.\nonumber\\ 
    & \left. -\frac{1}{2}h^k_0\partial_i\nabla^2h_{kj} + \frac{1}{2}h^k_0\partial_k\partial_l\partial_ih^l_j \frac{}{}\right) + 2\pi^{ij}\left(\frac{}{} \partial_ih_{0j}+K_{ij}\frac{}{}\right).
\end{align}
In this case, the structure of the primary constraints are the same of \eqref{PrimaryConstraints1} 
and the corresponding primary Hamiltonian is
\begin{equation}
\mathcal{H}_{1} = \mathcal{H}_{C} + \lambda_{\mu\nu} \phi^{\mu\nu} + \zeta_{ij}\psi^{ij} + \xi_{ij}\Sigma^{ij},
\end{equation}

where $\lambda_{\mu\nu}$, $\zeta_{ij}$ y $\xi_{ij}$ plays the same role as before. From  consistency on the primary constraints we identify the following  secondary constraints, they are  given by
\begin{equation}
    \begin{split}
        & S = \frac{1}{2}R_{ij} {}^{ij} +  \frac{2}{\mu}\epsilon^{ij}\partial_j\partial^kK_{ik} \approx 0, \\
        & S^{0i}  = \frac{\epsilon^{jk}}{4\mu} \partial_j \nabla^2 h^i_k - \frac{\epsilon^{jk}}{4\mu} \partial^i\partial_l\partial_jh^l_k + \partial_j\pi^{ij}\approx 0, \\
        & \overline{W} =  \frac{\epsilon^{ij}}{\mu}\partial_i\partial_lh^l_j + 2\pi \approx 0,
    \end{split}
\end{equation}
at this point, we can notice a subtle difference between this case and the $\lambda\neq\frac{1}{2}$ if we contrast the constraints $W$ and $\overline{W}$.
As consistency does not yield any new information, the process of generating constraints concludes. This leads us to the discovery of a set of 16 constraints
\begin{equation}
    \begin{split}
        & \phi^{00} : \pi^{00} \approx 0,\\
        & \phi^{0i} : \pi^{0i} \approx 0,\\
        & \phi^{ij} : \pi^{ij} - \alpha^{ij} \approx 0,\\
        & \psi^{ij} : \tau^{ij} \approx 0,\\
        & \Sigma^{11} : P^{11} - \frac{2}{\mu}K^1_2 \approx 0,\\
        & \Sigma^{12} : P^{12}-\frac{1}{2}(K^2_2-K^1_1) \approx 0,\\
        & \Sigma : P \approx 0,\\
        & S : \frac{1}{2}R_{ij}{}^{ij} + \frac{2}{\mu}\epsilon^{ij}\partial_j\partial^kK_{ik} \approx 0, \\
        & S^{0i} : \frac{\epsilon^{jk}}{4\mu} \partial_j \nabla^2 h^i_k - \frac{\epsilon^{jk}}{4\mu} \partial^i\partial_l\partial_jh^l_k + \partial_i\pi^{im} \approx 0, \\
        & \overline{W} : \frac{1}{\mu} \epsilon^{ij}\partial_i\partial_lh^l_j + 2\pi \approx 0.
    \end{split}
\end{equation}
In order to classify them into first-class and second-class constraints, we calculate the non-zero Poisson brackets between all constraints:
\begin{eqnarray}
\lbrace \phi^{ij},\psi^{kl} \rbrace &=& - \frac{1}{2}(\delta^{ik}\delta^{jl}+\delta^{il}\delta^{jk}) \delta^{2}(x-y),
\\
\lbrace \psi^{ij},S \rbrace &=& - \frac{1}{2}(\partial^{i}\partial^{j}-\delta^{ij}\nabla^{2}) \delta^{2}(x-y),
\\
\lbrace \phi^{ij},S^{0k} \rbrace &=& \frac{1}{8\mu}[\epsilon^{il}(\partial^{j}\partial^{k}+\delta^{jk}\nabla^{2}) + \epsilon^{jl}(\partial^{i}\partial^{k}+\delta^{ik}\nabla^{2})]\partial_{l} \delta^{2}(\vec{x}-\vec{y}),
\\
\lbrace \phi^{ij},\overline{W} \rbrace &=& \frac{1}{2\mu}(\epsilon^{ik}\partial^{j} + \epsilon^{jk}\partial^{i})\partial_{k} \delta^{2}(\vec{x}-\vec{y}),
\\
\lbrace \Sigma^{ij},\Sigma^{kl} \rbrace &=& - \frac{1}{\mu}(\epsilon^{ik}\delta^{jl} + \epsilon^{jk}\delta^{il} + \epsilon^{il}\delta^{jk} + \epsilon^{jl}\delta^{ik}) \delta^{2}(\vec{x}-\vec{y}),
\\
\lbrace \Sigma^{ij},S \rbrace &=& \frac{1}{\mu}(\epsilon^{ik}\partial^{j} + \epsilon^{jk}\partial^{i})\partial_{k} \delta^{2}(\vec{x}-\vec{y}),
\\
\lbrace S,\overline{W} \rbrace &=& \frac{1}{2}\nabla^{2}\delta^{2}(\vec{x}-\vec{y}).
\end{eqnarray}

These results lead us to the following 6 first-class constraints:
\begin{equation}\label{EH-CS: restricciones 1 class 2}
    \begin{split}
        & \Gamma_1 : \pi^{00} \approx 0,\\
        & \Gamma_2^{i} : \pi^{0i} \approx 0,\\
        &\Gamma_5 : P \approx 0,\\
        & \Gamma_4^{i} : \frac{\epsilon^{jk}}{4\mu} \partial_j \nabla^2 h^i_k - \frac{\epsilon^{jk}}{4\mu} \partial^i\partial_l\partial_jh^l_k + \partial_j\pi^{ij} \\
        & -\frac{\epsilon^{jk}}{4\mu}\partial_j\nabla^2\tau^i_k+ \frac{\epsilon^{jk}}{4\mu}\partial^i\partial_l\partial_j\tau^l_k \approx 0,
    \end{split}
\end{equation}
and to the following 10 second-class constraints:
\begin{equation}\label{EH-CS: restricciones 2 class}
    \begin{split}
        & \chi_1^{ij} : \pi^{ij} - \alpha^{ij} \approx 0,\\
        & \chi_2^{ij} : \tau^{ij} \approx 0,\\
        & \chi_3 : \frac{\epsilon^{ij}}{\mu} \partial_i \partial_l h^l_j + 2\pi \approx 0, \\
        & \chi_4 : P^{11} - \frac{2}{\mu}K^1_2 \approx 0,\\
        & \chi_5 : P^{12}-\frac{1}{\mu}(K^2_2-K^1_1) \approx 0,\\
        & \chi_6 : \frac{1}{2} R_{ij}{}^{ij} + \frac{2}{\mu} \epsilon^{ij} \partial_j \partial^kK_{ik} \approx 0. \\
    \end{split}
\end{equation}
By looking at equations \eqref{EH-CS: restricciones 1 class}-\eqref{EH-CS: restricciones 2 class1} and \eqref{EH-CS: restricciones 1 class 2}-\eqref{EH-CS: restricciones 2 class} we can recognize a similar structure in the constraints for $\lambda \neq \frac{1}{2}$ and $\lambda=\frac{1}{2}$ but they are strictly different, for the former  $P$ is a second-class constraint but is first-class for the later. Now, with the algebra of the second-class constraints (\ref{EH-CS: restricciones 2 class}) we can calculate  the matrix $C_{ab}=\left\lbrace \chi_{a}, \chi_{b}\right\rbrace$. Then, by using (\ref{Bdir}) again we find the following Dirac's brackets
\begin{eqnarray}
\label{DB2}
\{h_{ij}, \pi^{kl}\}_D&=& \frac{1}{2}\delta^{kl}_{ij} \delta^2(x-y) + \frac{\delta_{ij}}{4}\frac{\partial^k \partial^l}{ \nabla^2} \delta^2(x-y)+ \frac{\delta_{ij} \delta^{kl}} {4} \delta^2(x-y), \nonumber \\
\{K_{ij}, P^{kl}\}_D&=&  \frac{1}{2}\delta^{kl}_{ij} \delta^2(x-y)+ \frac{\delta^{1}_i\delta^1_j \delta^k_2 \delta_2^l }{2}\delta^2(x-y) - \frac{\delta^{1}_i\delta^1_j \delta^k_1 \delta_1^l }{2} \delta^2(x-y),        \nonumber \\
\{K_{ij}, \pi^{kl}\}_D&=&\frac{\delta^{11}_{ij} }{8\mu \nabla^2} \big\{ \epsilon^{mk} \partial_m \partial^l+  \epsilon^{ml} \partial_m \partial^k \Big\} (\partial_1^2- \partial_2^2) \delta^2(x-y), \nonumber \\
\{h_{ij}, K^{kl}\}_D&=& \frac{\delta_{ij} \delta^{11}_{kl}}{2 \nabla^2} (\partial_1^2-\partial_2^2) \delta^2(x-y)+ \frac{\delta_{ij} \delta^{12}_{kl} }{\nabla^2} \delta^2(x-y), \nonumber \\
\{K_{ij}, K^{kl}\}_D&=& \frac{\mu}{8} (\delta^{11}_{ij} \delta^{12}_{kl} - \delta^{12}_{ij} \delta^{11}_{kl}) \delta^2(x-y), \nonumber \\
\{P_{ij}, P^{kl}\}_D&=& \frac{\delta^{ij}_{12}}{4\mu} (\delta^{22}_{kl}- \delta^{11}_{kl})\delta^2(x-y)- \frac{\delta^{12}_{kl}}{4\mu} (\delta^{ij}_{22}- \delta^{ij}_{11}) \delta^2(x-y), \nonumber \\
\{h_{ij}, P^{kl}\}_D&=& \frac{\delta_{ij} \delta_{12}^{kl}}{\mu \nabla^2}(\partial_1^2-\partial_2^2) \delta^2(x-y) + \frac{\delta_{ij}}{\mu \nabla^2 } (\delta^{kl}_{22} -\delta_{11}^{kl}) \partial_1 \partial_2\delta^2(x-y)  \nonumber  \\
&+& \frac{2 \delta_{ij}}{\mu \nabla^2}\big(  \epsilon^{km} \partial_m \partial^l+  \epsilon^{lm} \partial_m \partial^k  \big) \delta^2(x-y). 
\end{eqnarray} 
Where we can observe that the brackets (\ref{DB1}) and (\ref{DB2}) are different from each other. Thus, this is a sign that the theories are not equivalent. It is worth commenting that in the $\lambda R$ gravity, the contribution of $\lambda$ is irrelevant. However, in the theory under study, the difference between these  models for different values of $\lambda$ is manifested. \\
In addition, once the constraints have been classified,  the counting of physical degrees of freedom is performed as follows: there are 24 canonical variables, six first-class constraints and 10 second-class constraints, therefore 
\begin{equation}
DoF = \frac{1}{2}\left( 24 - 2*6-10 \right) = 1, 
\end{equation}
thus, for different values of $\lambda$ we identify one physical degree of freedom. Nevertheless, the canonical structure  is different.

\section{Conclusions}

The Hamiltonian analysis of the $\lambda R$ model plus the  Chern-Simons theory has been carried out. The identification of all constraints was performed directly. We showed that the constraint structure of the theory is different for both cases of $\lambda$; in particular, the constraint $P$ is first-class for $\lambda\neq\frac{1}{2}$ while it is second class for $\lambda=\frac{1}{2}$; for the former, this means  that the gauge transformations generated by $P$ on $K_{ij}$ are $K_{ij} \rightarrow K_{ij} + \eta_{ij}\epsilon$, for the later this is not a symmetry for theory.   Moreover, if we observe the constraints of the case $\lambda\neq\frac{1}{2}$, the parameter $\lambda$ appears directly in the set of first-class constraints $\Gamma_{3}$ and in the second-class constraint $\chi_{6}$, thus manifesting the deviation concerning topologically massive gravity. Despite the difference in the constraints between the selected cases of $\lambda$, the number of degrees of freedom remains the same.\\
Furthermore, the second-class constraints were removed successfully, and we calculated the corresponding Dirac brackets. The deviation remains in the canonical structure; at the end, one can recover the structure of topologically massive gravity  if the limit $\lambda\rightarrow1$ is taken.\\
We have added an appendix to discuss the equations of motion and the gauge fixing for the final structure of Dirac's brackets. We can observe that the $\lambda$ parameter is relevant in that structure, most at this classical level. Of course, it is necessary to calculate the propagators of the fields to observe the physical relevance of the $\lambda$ parameter in the quantum scenary. However, this point is still in progress, and we expect to report new results soon.   

\section{Appendix: A note on degrees of freedom}

It is known that when the Chern-Simons action is coupled to GR in $2+1$ dimensions, an extra degree of freedom emerges. In the model considered in this paper, something similar happens because a single degree of freedom arises. We can say more about this degree of freedom. In fact,  let's consider the equations of motion; since the introduction of the $\lambda$ parameter in the standard kinetic requires a $2+1$ decomposition, we need to calculate the equations of motion for the fields. A long but straightforward calculation of the Euler-Lagrange equations of \eqref{L1} yields
\begin{eqnarray}
\label{EoMh00}
\square\Big{[}h_{00} + \frac{2}{\mu}\epsilon^{ij}\partial_{j}h_{0i} \Big{]} &=& 0,
\\
\label{EoMh0i}
\square\Big{[} h_{0}{}^{i}  - \frac{1}{\mu}[\epsilon^{jk}\partial_{j}h^{i}{}_{k} + \epsilon^{ij}(\dot{h}_{0j} - \partial_{j}h_{00})] \Big{]} + (\lambda - 1)\partial^{i}\dot{h}_{00} &=& 0,
\\
\label{EoMhij}
\square\Big[h^{ij}  + \frac{1}{\mu}[\epsilon^{ik}(\partial_{k}h_{0j}-\dot{h}^{j}{}_{k}) + \epsilon^{jk}(\partial_{k}h_{0}{}^{i}-\dot{h}^{i}{}_{k})]\Big{]} + (\lambda-1)\delta^{ij}\ddot{h}_{00} &=& 0,
\end{eqnarray}
where we have taken  into account the Coulomb gauge and the traceless condition. On the other hand, the equations of motion of topologically massive gravity under the same conditions reads:
\begin{eqnarray}
\label{EoMTMG}
\square\Big{[} h_{\alpha\beta} + \frac{1}{\mu}(\epsilon_{\alpha}{}^{\rho\nu}\partial_{\rho}h_{\nu\beta} + \epsilon_{\beta}{}^{\rho\nu}\partial_{\rho}h_{\nu\alpha}) \Big{]} = 0
\end{eqnarray}
this expression can be reduced to a generic Klein-Gordon equation for $h_{\mu \nu}$. The components of (50) correspond to the d’Alembertian terms of (47)-(49). Nevertheless, in equations (48) and (49), an extra term is present due to the modification of the kinetic term; at this stage, this can be considered a deviation from topologically massive gravity.
\

Now, if we choose the following gauge
\begin{eqnarray}
\label{GF: h00}
h_{00} \approx 0,
\\
h_{0i} \approx 0,
\\
\delta^{ij}K_{ij} \approx 0,
\\
\partial_{j}h^{ij} \approx 0,
\end{eqnarray}
all first-class constraints become second-class constraints:
\begin{eqnarray}
\Phi_{1} &=& h_{00}  \approx 0,
\\
\Phi_{2} &=& \pi^{00}\approx 0,
\\
\Phi_{3i} &=& h_{0i}\approx 0,
\\
\Phi_{4}^{i}&=& \pi^{0i}\approx 0,
\\
\Phi_{5} &=& \frac{1}{2}(\partial^{i}\partial^{j} - \delta^{ij}\nabla^{2})h_{ij} + \frac{2}{\mu}\epsilon^{ij}\partial_{j}\partial^{k}K_{ik} \approx 0,
\\
\Phi_{6} &=& \delta^{ij}K_{ij}\approx 0,
\\
\Phi_{7}^{i} &=& \partial_{j}\pi^{ij} + \frac{1}{4\mu}\epsilon^{jk}(\partial_{j}\nabla^{2}h^{i}{}_{k} - \partial^{i}\partial_{j}\partial^{l}h_{kl}) \approx 0,
\\
\Phi_{8}^{i} &=& \partial_{j}h^{ij}\approx 0.
\end{eqnarray}
The contribution of the parameter $\lambda$ occurs  in the brackets structure, for example, for the pairs $(h_{ij},K^{kl})$ and $(K_{ij},\pi^{kl})$ we find
\begin{eqnarray}
\lbrace h_{ij},K^{kl} \rbrace_{D'} &=& - \Big{[}\delta_{ij}+\frac{\partial_{i}\partial_{j}}{\nabla^{2}}\Big{]}\Big{[}\Big{(}\frac{2}{1-2\lambda} + 1\Big{)}\delta^{kl} - 2\frac{\partial^{k}\partial^{l}}{\nabla^{2}} \Big{]} \delta^{2}(x-y),
\\
\lbrace K_{ij},\pi^{kl} \rbrace_{D'} &=& - \frac{1}{4\mu}\Big{[}2\Big{(}\frac{2}{1-2\lambda}+1\Big{)}\delta_{ij} - 2\frac{\partial_{i}\partial_{j}}{\nabla^{2}}\Big{]} \Big{[}\epsilon^{mk}\partial^{l}+\epsilon^{ml}\partial^{k}\Big{]}\partial_{m}\delta^{2}(x-y),
\end{eqnarray}
if we take the value $\lambda=1$, then we recover the brackets reported in \cite{Barcelos, Esca}. At the end, by virtue of eq. \eqref{GF: h00} we can discard the corresponding terms of $h_{00}$ and we recover a Klein-Gordon equation for the field $h_{\mu \nu}$.




\end{document}